\documentclass[prl,twocolumn,showpacs]{revtex4}

\usepackage{amssymb,amsmath,bm,epsfig,graphicx,subfigure}

\pdfoutput=1

\begin{document}

\title{Charge and spin transport on graphene grain boundaries \\ in a quantizing magnetic field}
\author{Madeleine Phillips and E. J. Mele}
    \email{mele@physics.upenn.edu}
    \affiliation{Department of Physics and Astronomy, University of Pennsylvania, Philadelphia PA 19104  \\}
\date{\today}

\begin{abstract}
We study charge and spin transport along grain boundaries in single layer graphene in the presence of a quantizing perpendicular magnetic field.  Transport states in a grain boundary are produced by hybridization of Landau zero modes with interfacial states.  In selected energy regimes quantum Hall edge states can be deflected either fully or partially into grain boundary states. The degree of edge state deflection is studied in the nonlocal conductance and in the shot noise. We also consider the possibility of grain boundaries as gate-switchable spin filters, a functionality enabled by counterpropagating transport channels laterally confined in the grain boundary.
\end{abstract}

 \maketitle

Grain boundaries in single layer graphene (SLG) can host laterally confined electronic states inside interface-projected bandgaps \cite{Mesaros,YL1,MP_GM,Mayer}. They have attracted attention as possible linear transport channels (i.e. as atomically defined nanowires) \cite{Lahiri}, but any practical implementation is limited by generally small group velocities along the boundary line and the difficulty that they are embedded in graphene which is a conductive medium. In this Letter we study an approach that avoids both difficulties and supports robust atomically defined gate-switchable transport channels in SLG grain boundaries.

Our approach exploits a unique feature of dispersive edge states that derives from the SLG zero modes in a uniform magnetic field. The zero modes are the signature of Landau quantization in a massless relativistic system.  Crucially they are polarized on complementary sublattices in two independent ${\mathbf{k}}$-space valleys and they do not develop into dispersive edge modes by soft electrostatic confinement on an outer boundary. Instead the lowest Landau level becomes dispersive at an atomically defined edge by hybridizing with the surface states to form {\it pairs} of separately dispersive particle and hole branches \cite{Brey}. Here we show that on interior channels defined by twin grain boundaries (t-GB's) a similar mechanism generates propagating nonchiral modes confined to the grain boundary and within the bulk Landau gaps.  These modes can serve as interior transport channels in which backscattering is suppressed by crystal momentum conservation. Further, they are embedded in an incompressible electronic medium, and we demonstrate that this allows gate-switchable one dimensional interior transport pathways which can be readily accessed from the quantum Hall edge states on an outer boundary.

Previous work on GB's in SLG but without a magnetic field has studied their spectra of laterally confined electronic states \cite{Mesaros,YL1,Mayer} and the effect of the misorientation angle between neighboring grains on ballistic transport perpendicular to the boundary \cite{YazyevLouie}. In this Letter we focus instead on the possibility of directed transport {\it along} a GB and show that in the quantum Hall regime the hybridization of Landau levels with GB states produces dispersive transport states on the wall.  This provides inter-edge transport channels that break the quantization of the Hall conductance in agreement with earlier work \cite{Bergvall, Cummings, DalLago, Lafont}.  Here we explore some unique functionalities enabled by this mechanism and find that it supports configurable interior transport channels that can be switched ``on"  with an electostatic gate and allow spin filtering with high fidelity over a tunable energy range.
\begin{figure}
  \includegraphics[angle=0,width=0.8\columnwidth]{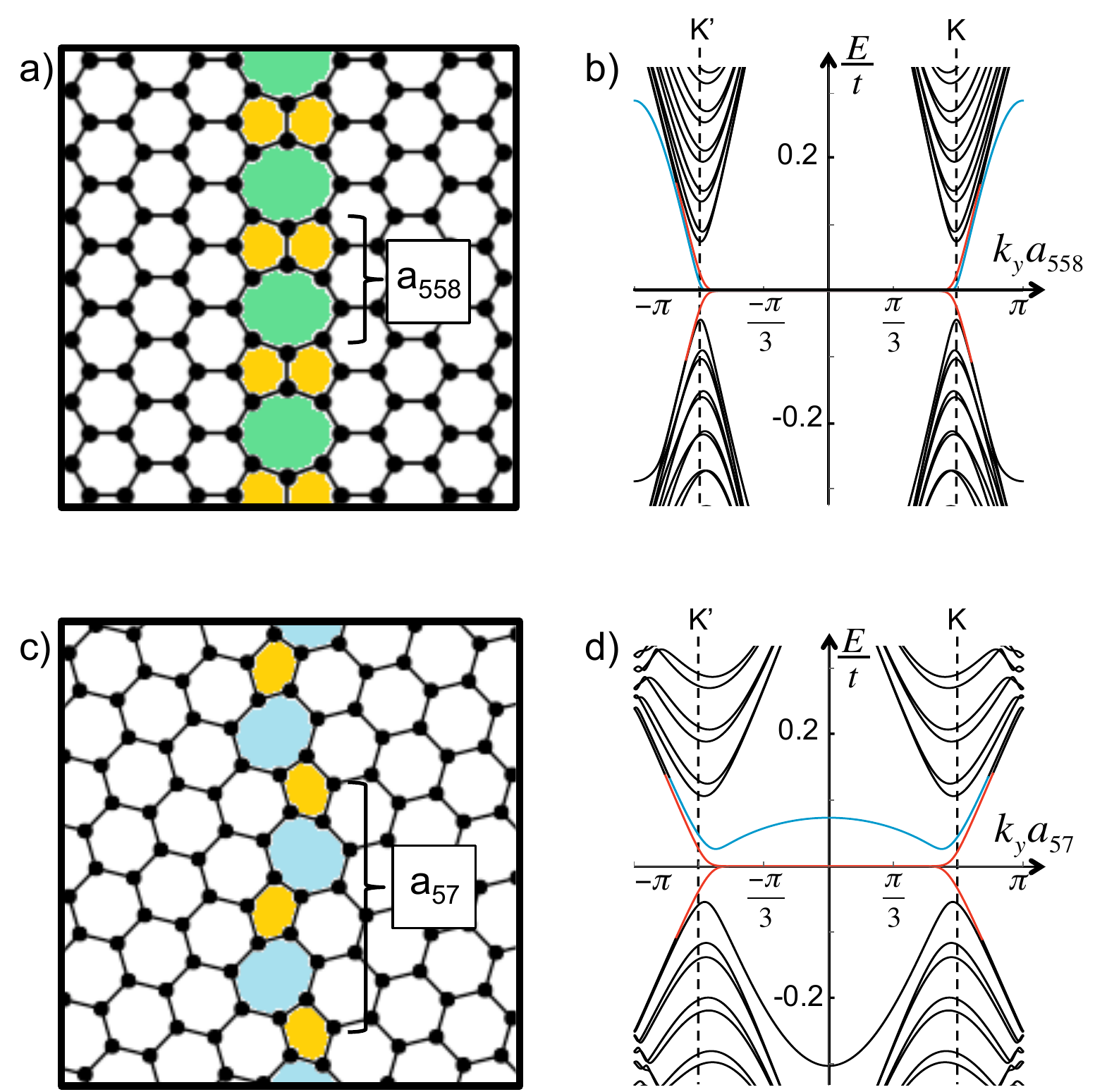}
  \caption{\label{lattices_0Bands} Grain boundary lattices and their band structures in zero magnetic field.  {\bf a)} The 5-5-8 t-GB structure with period $a_{558}=2a$, where $a$ is the graphene lattice constant. {\bf b)} Tight-binding band structure in units of the nearest neighbor tight binding parameter $t$ for a 5-5-8 ribbon bicrystal with its low energy surface (interfacial) bands highlighted in red (blue). The flat band is triply degenerate with two surface states on opposite edges and one interfacial state. {\bf c)} The 5-7 t-GB structure with period $a_{57}=\sqrt{13}a$. {\bf d)} Band structure for a 5-7 ribbon bicrystal.  Flat bands are doubly degenerate surface bands. }
\end{figure}

To illustrate these effects we present results for representative t-GB geometries: the 5-5-8 GB (Fig. \ref{lattices_0Bands}a) a periodic line of octagons and pentagon pairs and the 5-7 GB (Fig. \ref{lattices_0Bands}c) a line of pentagons and heptagons. The transport properties we obtain for these structures derive from their lattice connectivities and are described most clearly within a nearest neighbor tight binding Hamiltonian that ignores effects of small out of plane structural relaxations. The band structures for ribbon bicrystals partitioned by these GB's but without a magnetic field (Fig. \ref{lattices_0Bands}b, d) show the closure of their projected bulk band gaps at critical values of the parallel momentum $k_y$ and the appearance of surface/interface states within the gaps at low energy. For the case of the 5-5-8 t-GB the interfacial state is a symmetry protected flat band at $E=0$ degenerate with the outer edge surface bands \cite{MP_GM}. The 5-7 structure confines a narrow band of interfacial states away from the charge neutrality point as well as flat band surface states.

\begin{figure}
  \includegraphics[angle=0,width=0.8\columnwidth]{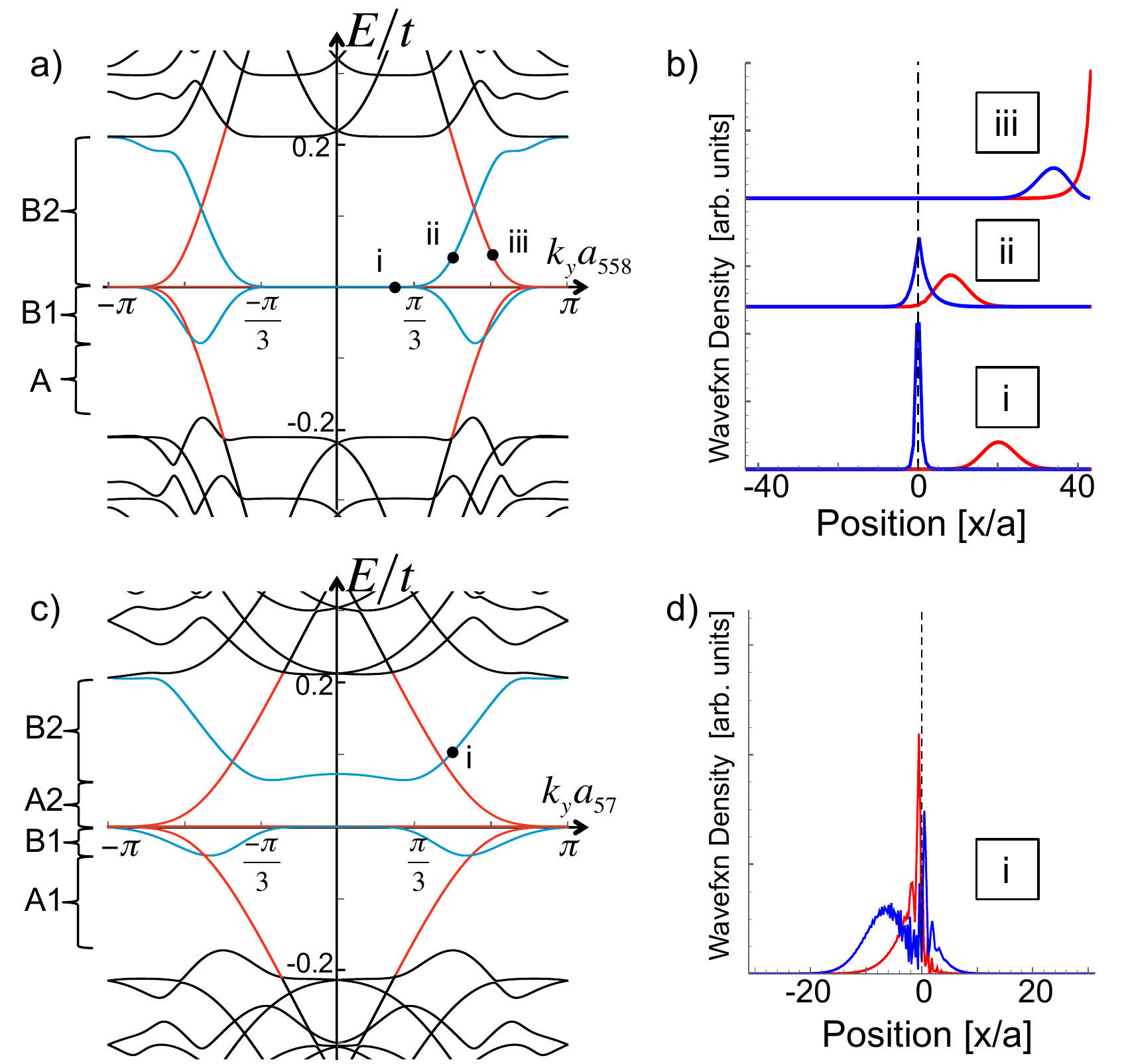}
  \caption{\label{BandsB} Band structures for {\bf a)} 5-5-8 and {\bf c)} 5-7 t-GBs in a perpendicular magnetic field. Black bands correspond to bulk states.  Surface (interfacial) bands are highlighted in red (blue). Energy regions where only chiral edge modes exist are marked A, while energy regions where both edge and interfacial transport modes exist are marked B.  Flat bands near $k_ya=0$ are triply (doubly) degenerate for 5-5-8 (5-7).  The right panel shows charge densities projected onto the $A$ (red) and $B$ (blue) sublattices for energies/momenta marked in the band structures for {\bf b)} the 5-5-8 and {\bf d)} the 5-7 t-GB.  State i in panel {\bf b)} corresponds to one of three degenerate states at that $k_y$ value.  The overlap of surface and interfacial states with LLL states generates dispersing modes.}
\end{figure}

We focus on the pattern of edge and interfacial modes that occur when the bulk is gapped by a quantizing magnetic field.  In the presence of a perpendicular field $B$ we couple electronic motion to a vector potential $\vec{A} = Bx \, \hat{y}$ which retains periodicity along the GB and consider a field strength for which magnetic length, $\sqrt{\hbar/eB} \ll w$ where $w$ is the ribbon width. Note that the bulk lowest Landau level (LLL) is polarized on opposite sublattices in each valley and consequently it is pinned at $E=0$. By constrast at momenta $k_y$ where the LLL wavefunction overlaps a grain boundary line it hybridizes with interfacial states to form a pair of composite dispersive bands.  This is shown in the band structures and in the plots of the charge densities projected onto the $A$ and $B$ sublattices in Fig. \ref{BandsB}.
In the band structures, the outer edges modes (red) are chiral one-way modes with opposite velocities on opposite edges.  This occurs because of the mismatch of the Chern number assigned to the first Landau gap (${\cal C}=\pm 1$) and to the vacuum (${\cal C} =0$) which protects one chiral branch on each outer edge. By contrast across the interior grain boundary  $\Delta {\cal C}=0$ and  interfacial modes appear only in dispersing pairs with opposite velocities on the same boundary. These modes are the two branches of a one dimensional band with its ``Fermi points"  displaced in momenta. This can be contrasted with the situation for SLG snake states on interior boundaries which can be produced in the quantum Hall regime either by reversal of the magnetic field or by sign reversal of the carrier type in a uniform field \cite{Rickhaus, SnakeStates} where $\Delta {\cal C} = \pm 2$ protects pairs of co-propagating chiral modes on the domain wall.  Backscattering is prohibited for snake states while for the GB states it is suppressed for smooth impurity potentials.

We consider a scenario where an outer quantum Hall edge state can be fully or partially deflected into the grain boundary. This type of deflection is generally very weak in an isolated graphene flake. Edge states and domain wall states contact only at isolated points of intersection.  If the coupling between these channels were zero at these intersections then the momenta of the two degrees of freedom would be separately quantized  in units $\propto 1/{\cal L}_c$ and $\propto 1/{\cal L}_w$ respectively where ${\cal L}_c$ is the flake perimeter and ${\cal L}_w$ is the length of the domain wall.  Short range matrix elements connecting these two degrees of freedom scale  $\propto 1/\sqrt{{\cal L}_c {\cal L}_w}$ and therefore couple these channels effectively only when the mixing scale greatly exceeds their energy separation, i.e. when $\sqrt{{\cal L}_w/{\cal L}_c} \ll 1$ which is typically not satisfied on a graphene flake. We have confirmed this kinematical obstruction in calculations on large flakes partitioned by grain boundaries where we find that edge modes and GB modes coexist as (nearly) independent degrees of freedom within the bulk Landau gaps. The former states circulate on the outer boundary and the latter states are observed as Fabry-Perot-like standing wave states laterally confined to the GB. The degree of decoupling is evident in the spatial distributions of the charge densities and velocity fields calculated for these states \cite{Supp}. States confined to the GB can be understood as the modes of a resonant cavity with a very high reflectivity at the contacts to the outer edges which does not admit useful switching of GB current pathways to or from an outer edge.

\begin{figure}
  \includegraphics[angle=0,width=0.9\columnwidth]{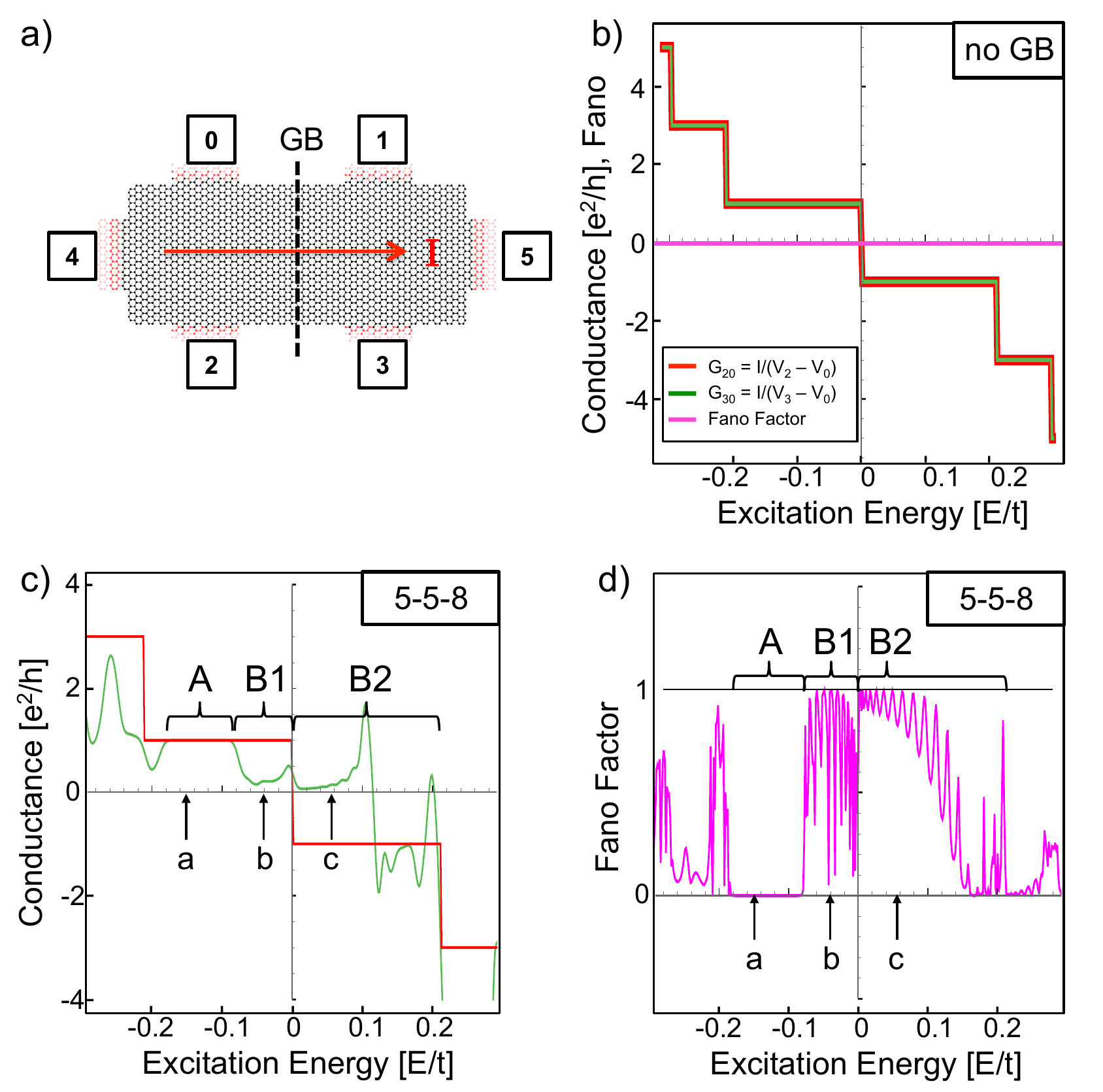}
  \caption{\label{Gs_Fanos} {\bf a)} Hall bar geometry for calculating conductances.  Current $I$ is injected at lead 4 and collected at lead 5.  Voltages are calculated at each lead. {\bf b)} Conductances measured from lead 2 to lead 0 ($G_{20}$, green curve) and from lead 3 to lead 0 ($G_{30}$, red curve) for a Hall bar with no GB and Fano factor (pink curve) for a two-terminal device. {\bf c)} Conductances $G_{20}$ (red) and $G_{30}$ (green) for a Hall bar partitioned by a 5-5-8 GB.  In the 1st LL gap, Hall conductance plateaus are destroyed in the B regions and maintained in the A regions. (See Fig. \ref{BandsB}a.) ($G_{30}$ values are Gaussian averaged with a peak width of $c=0.005 E/t$.)  {\bf d)} Fano factor ($F$) for a two-terminal device partitioned by a 5-5-8 GB showing the transition from $F=0$ (no deflection) to $F \sim 1$ (large deflection).  Oscillations in $F$ are due to the finite length of the GB. Energies marked a, b, c in {\bf c)} and {\bf d)} are identified with current maps in Figure 4a, b, c. }
\end{figure}

\begin{figure}
  \includegraphics[angle=0,width=0.8\columnwidth]{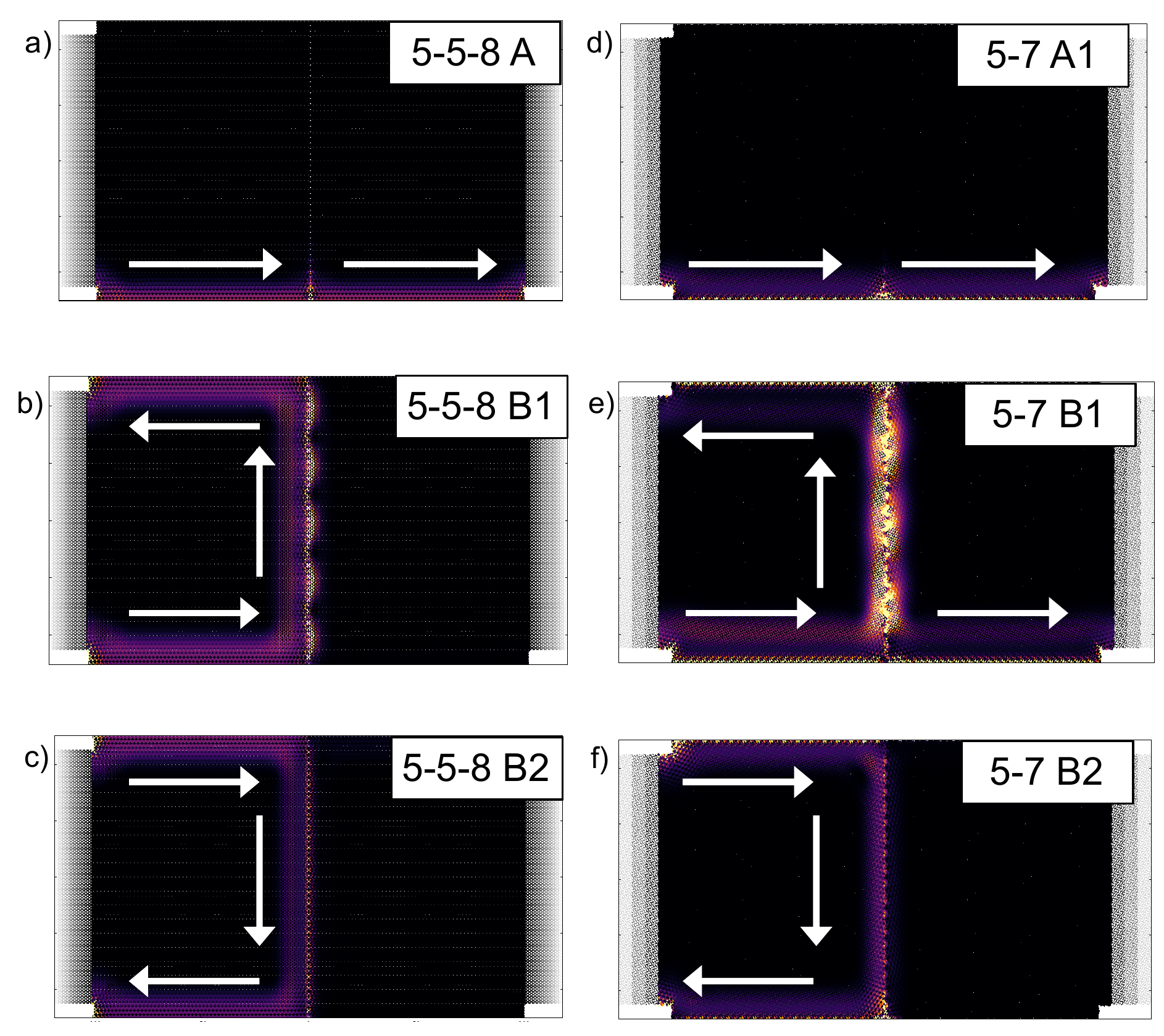}
  \caption{\label{States} Maps of current distribution originating from the left (source) lead. All devices are $200a$ x $100a$ with lead width $80a$, where $a=0.246nm$ is the graphene lattice constant.  {\bf a), b), c)} Currents in the 5-5-8 GB system at energies marked by arrows a, b, c in Figure \ref{Gs_Fanos}c and \ref{Gs_Fanos}d. {\bf d), e), f)} Currents in the 5-7 GB system in the A1, B1, and B2 regimes. Currents in the A spectral regions ({\bf a, d}) are confined to the sample edges and in the B spectral regions ({\bf b, c, e, f}) have strong deflection into the interfacial transport states. }
\end{figure}

This kinematic decoupling can be avoided in an open system where the graphene is in contact with external particle reservoirs and its electronic states are lifetime broadened.  We study transport in this regime using the scattering matrix formalism implemented in Kwant \cite{Kwant} to calculate the nonlocal conductances of multi-terminal Hall bars (Fig. \ref{Gs_Fanos}). We consider a current biased system with current injected at electrode 4 and collected at 5 and obtain the nonlocal conductances by calculating the voltage states on all electrodes. Conductances calculated for a Hall bar without a grain boundary show quantized plateaus in the (spinless) Hall conductance at values $(2n+1) e^2/h$ (Fig. \ref{Gs_Fanos}b). When the Hall bar is partitioned by a GB the Hall plateaus are fully supported in a spectral range (marked A in Fig. \ref{Gs_Fanos}c) where only states on the outer edges are accessible.  Outside this energy window  but still within the bulk Landau gap (B regions), the system supports transport states both on the edge and along the grain boundary. In this case transverse conductance measured between leads on the same side of the grain boundary (red curve) retains quantized Hall plateaus whereas the conductance between contacts that straddle the GB (green curve) is reduced as the system develops a nonzero longitudinal potential difference along a common edge.
As seen in Figure \ref{States}, the A energy regions where Hall plateaus are preserved correspond with transport only along the sample edges, while the B regions correspond to edge currents that can be efficiently deflected into the GB.  Tuning the chemical potential accesses these two distinct energy regimes within the Landau gap.

In the B spectral window, outer edge channels can be nearly completely deflected through GB transport states and into the opposite edges of the Hall bar, effectively short circuiting the quantum Hall effect (Fig. \ref{States}b, c, f).  These states are associated with regions where the cross-GB conductance (green curve, Fig. \ref{Gs_Fanos}c) is nearly zero, indicating a voltage drop on the same edge arising from a large contact resistance at the GB/edge intersection at these energies.  The partitioning of the current at the GB can be further quantified by studying the two terminal shot noise and Fano factor which gives the ratio of the shot noise to its value in the uncorrelated (Poisson) limit. The Fano factor is $F=\sum_{n} T_{n}(1-T_{n})/\sum_{n} T_{n}$, where $T_n$ is the transmission probability in the $n-th$ channel \cite{Fano}.  Importantly, $F$ is a good measure of the degree of edge state deflection into the GB because the chiral edge states cannot backscatter. Thus $F=0$ is the signature of perfect transmission from source to drain through an edge state. Conversely $F \sim 1$ occurs only in the strong deflection regime where there are weak residual correlations between events that transmit discrete charges to the drain.  Figure \ref{Gs_Fanos}d shows that the Fano factor as a function of energy displays a sharp crossover from the edge state ($F=0$) to the interior channel ($F \sim 1$) regimes.

\begin{figure}
  \includegraphics[angle=0,width=0.75\columnwidth]{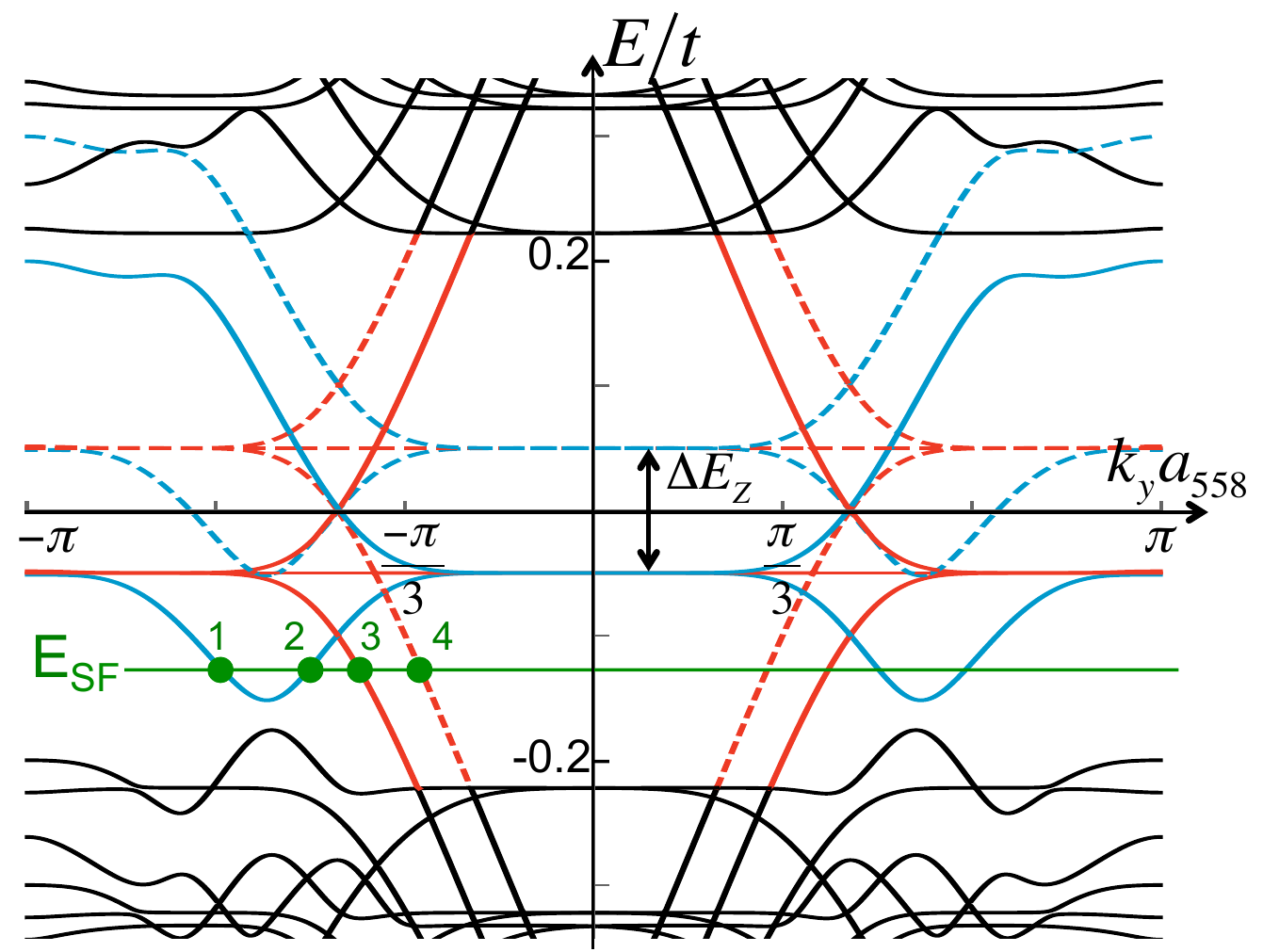}
  \caption{\label{558wSpin} Band structure for 5-5-8 t-GB ribbon with Zeeman splitting $\Delta E_{Z} = g\mu_{B}B$. Solid (dashed) bands correspond to spin up (down) states and red (blue) bands are edge (GB) states.  The GB-edge contact spin-filters the deflection of edge transport states from edge channels 3,4 to GB channel 1. }
\end{figure}

These junctions can also be used as gate switchable spin filters. To study this we re-introduce the Zeeman coupling of the electron spin to the magnetic field. As shown in Figure \ref{558wSpin} the spectrum is then spin split, and over a range of energy $\sim g \mu_B B$ the edge mode can be deflected to the interior channel only for a single spin polarization. Note that this construction yields a spatial spin filter (i.e. it is a spin selective beam splitter) using the deflection into a GB current pathway only for a single spin polarization. This can be compared with an earlier proposal for spin-selective transport at the edges of SLG in the quantum Hall regime \cite{Abanin,Young} where the Zeeman splitting of the Landau zero modes allows the particle and hole branches of its edge state spectrum to overlap in energy and produce helical edge modes.  The helical modes are branches with opposite spin polarizations and velocities but residing on a common spatial boundary in contrast to the spatial separation of spin currents that occur for deflection into a GB.

These transport effects are seen most clearly in the first Landau gap. At higher energies edge states and grain boundary states also couple but the spectra are congested as interfacial states begin to mix more with higher LL states, and the fidelity of the deflection is reduced.

Experimental demonstration of inter-edge charge transport along a graphene GB appears to be well within reach of currently available materials.  The fabrication of SLG containing 5-5-8 GB's  has been demonstrated \cite{Lahiri, Chen} and the relevant energy ranges are accessible. For the pseudorelativistic spectrum in graphene the fundamental Landau gap between the zeroth and the first Landau levels is $\Delta_1 \sim 65 \, {\rm K} \, \sqrt{B(T)}$, providing a sizeable energy window for observing these effects.  The conditions for spin filtering are more restrictive since the bare Zeeman splitting $\Delta_z = g \mu_B B \sim 1.5 \,  {\rm K} \,  B(T)$ ($g=2$) before accounting for an exchange enhancement of the $g$-factor \cite{Abanin,Volkov}. Note that strong interactions further complicate the situation in graphene in a strong perpendicular magnetic field since they gap the bulk even at charge neutrality \cite{Abanin,Young}. Although this has been attributed to an interaction-driven spin {\it unpolarized} state, experiments find that this phase undergoes a closing of the bulk gap marking a transition to a spin polarized state at only moderate value of the parallel magnetic field, $B_{\|} \sim 20 \, T$ {\cite{Young}.  Transport measurements on the edge in this latter state show the signature of ballistic motion in helical channels within an energy window $g \mu_B B_{\rm tot}$. This high (tipped) field Zeeman-dominated phase is a promising setting for the spin filtering along a GB found in our calculations.

This work was supported by the Department of Energy under grant DE-FG02-84ER45118.

\pagebreak

\widetext
\begin{center}
\textbf{\large Supplemental Materials: Charge and spin transport \\ on graphene grain boundaries in a quantizing magnetic field}

\medskip
\medskip
\medskip
Madeleine Phillips and E. J. Mele*

\textit{Department of Physics and Astronomy,\\ University of Pennsylvania, Philadelphia PA 19104}
\end{center}

\setcounter{equation}{0}
\setcounter{figure}{0}
\setcounter{table}{0}
\setcounter{page}{1}
\makeatletter
\renewcommand{\theequation}{S\arabic{equation}}
\renewcommand{\thefigure}{S\arabic{figure}}
\renewcommand{\bibnumfmt}[1]{[S#1]}
\renewcommand{\citenumfont}[1]{S#1}

\section{Isolated Flake Calculations}
To study current patterns on isolated graphene flakes partitioned by grain boundaries, we write down a nearest neighbor tight binding Hamiltonian with a perpendicular magnetic field introduced via the Peierls substitution:
\begin{equation}
H = \sum_{<m, n>} te^{i\frac{e}{\hbar}\int_{r_n}^{r_m}{\bf A}\cdot{\bf dl}}c^{\dag}_{m}c_{n}
\end{equation}
where $t=-1$ is the hopping parameter, and the sum is over nearest neighbors $m$ and $n$.  ${\bf A} = Bx{\bf \hat{y}}$ is the Landau gauge vector potential for a perpendicular magnetic field.  To plot the charge density on the flake (Fig. \ref{Dens_Vort}a, c), we diagonalize the Hamiltonian and plot the wavefuntion amplitude for a given energy, $|\psi_E|^2$.

\begin{figure}
  \includegraphics[angle=0,width=1.0\columnwidth]{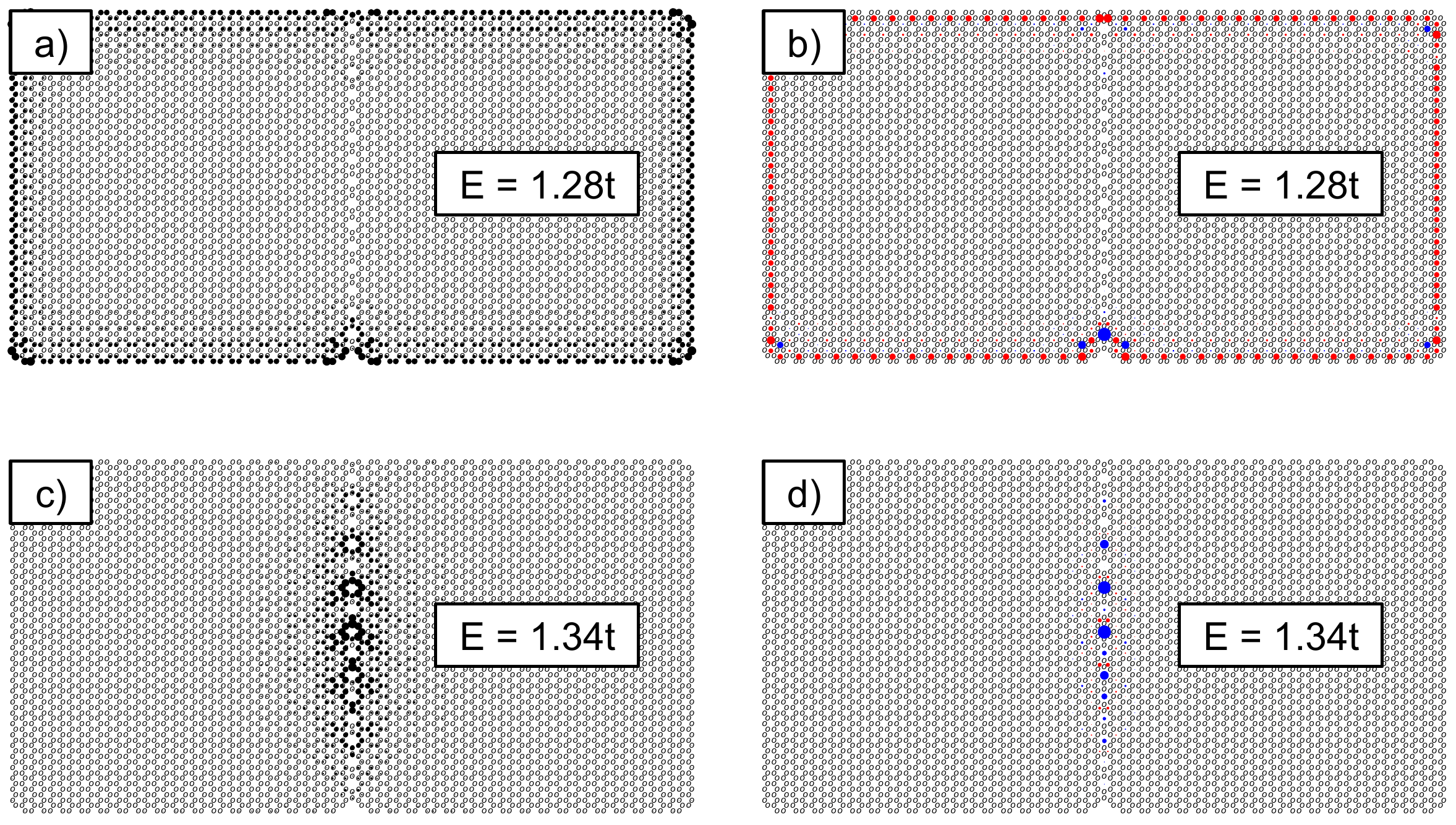}
  \caption{\label{Dens_Vort} The left panel shows charge densities for the 5-5-8 t-GB at {\bf a)} $E=1.28t$ and {\bf c)} $E=1.34t$.  Open circles are graphene lattice sites.  Filled circles represent wavefunction amplitude squared at a given site, with the radius of the circles denoting the relative magnitude of the density at each site. The right panel shows circulation of the velocity field around each plaquette for the 5-5-8 t-GB at {\bf b)} $E=1.28t$ and {\bf d)} $E=1.34t$.  Red circles represent net clockwise circulation, and blue circles represent net counter clockwise circulation.  The radius of each filled circle denotes the relative magnitude of the net circulation. The size of the flake in each panel is $64a\times32a$, where $a=0.246nm$ is the graphene lattice constant.  All calculations are carried out in a perpendicular magnetic field pointing out of the plane such that magnetic flux through a hexagonal plaquette is $\Phi=1/3$.  $E=1.28t$ and $E=1.34t$ are both in a $C=1$ gap (where $C$ is the Chern number) of the Hofstadter butterfly that is adiabatically connected with the first Landau level gap at $\Phi \ll 1$.  The state imaged in {\bf a, b)} is a chiral edge state due to the quantum Hall effect.  In contrast, the state plotted in {\bf c, d)} has a standing wave pattern arising from backscattering, indicating that the state localized on the GB is not chiral. Despite the small energy separation of the edge and GB states shown ($\Delta E = 0.06t$), there is generically no coupling between them, indicating that in the isolated flake, GB's do not ``short-circuit" the quantum Hall edge currents. }
\end{figure}

To visualize the current, we plot the circulation of the velocity field (Fig. \ref{Dens_Vort}b, d).  The net velocity operator on a given bond from site $j$ to site $i$ is given by the current on that bond multiplied by the real-space vector pointing from $j$ to $i$:
\begin{equation}
{\bf \hat{v}}_{ji} = \frac{-i}{\hbar}({\bf r}_i - {\bf r}_j) (t_{ij}c^\dag_ic_j - t^*_{ij}c^\dag_jc_i)
\end{equation}
where $c_j$ ($c^\dag_j$) annihilates (creates) an electron at site $j$.  The hopping amplitude from $j$ to $i$, $t_{ij}$, includes the Peierls phase, and $t^*_{ij}=t_{ji}$.  Because we include only nearest-neighbor hopping, each vector in the velocity field is directed along a bond in the lattice network.  It is natural to compute the circulation of the velocity field around closed loops defined by plaquette boundaries.  We obtain the circulation of the velocity, $C$, for each plaquette, $P$, by calculating
\begin{equation}
C_P = \oint_{\partial P} {\bf v}\cdot {\bf dl}
\end{equation}
The circulation can equivalently be expressed as the integral of the vorticity, ${\boldsymbol \omega} = \nabla \times {\bf v}$, over the plaquette area. When there is net counter-clockwise current around a plaquette, $C_P>0$, and when there is net clockwise current around a plaquette, $C_P<0$.  In Figure \ref{Dens_Vort}, we plot positive circulation with a blue dot at the center of the corresponding plaquette and negative circulation with a red dot where the radius of the dot signifies the relative magnitude of the circulation.  Thus, a uniform clockwise current around the sample is represented by red dots of constant radius around the perimeter of the flake (Fig. \ref{Dens_Vort}b), which is the expected quantum Hall edge mode for an electron current produced by a magnetic field pointing out of the plane.  This edge current configuration is separated from the standing wave current bound on the GB (Fig. \ref{Dens_Vort}d) by only $\Delta E = 0.06t$.  In the isolated flake with no peak broadening due to coupling with reservoirs, even this small energy separation is enough to decouple grain boundary states from edge states, which prevents edge current deflection for generic geometries.

\section{Open System Calculations in Kwant}
The scattering matrix calculations for a GB-partitioned flake coupled to semi-infinite leads are carried out using the Kwant package for python \cite{S_Kwant}.  The Hamiltonian for both scattering region and leads is again a nearest neighbor tight-binding model with a perpendicular magnetic field introduced everywhere via a Peierls substitution.  The vector potential in the scattering region and the vertical leads is ${\bf A} = Bx {\bf \hat{y}}$ as before, but the vector potential in the horizontal leads is ${\bf A} = -By{\bf \hat{x}}$ such that the lead is fully periodic along ${\bf \hat{x}}$ and allows for plane waves traveling along the lead.  Hexagonal plaquettes at the interface between ${\bf A}$ regions have  Peierls phases on each bond chosen to ensure that all hexagonal plaquettes have equal magnetic flux.

The samples used in the Kwant conductance calculations are six terminal Hall bars partitioned by a GB perpendicular to the source/drain leads.  The S-matrix for a given geometry is calculated in Kwant and then used to obtain the conductance matrix, ${\bf G}$, via the Landauer-B\"{u}ttiker formalism.  We then solve for the voltages at each lead:
\begin{equation}
{\bf V} = {\bf G}^{-1} {\bf I}
\end{equation}
where ${\bf I}$ is a vector that takes values of $\pm I$ at the source/drain lead and zero at all other leads.  Using these voltages, we compute the nonlocal conductance measured between leads $x$ and $y$ from the expression $G_{xy}=I/(V_x - V_y)$.  The Fano factor is computed on two-terminal systems using the equation given in the main text, where the transmission probabilities are calculated from the S-matrix.   Both non-local conductances and Fano factors are plotted in Fig. \ref{Gs_Fs_Sup}.

\begin{figure}
  \includegraphics[angle=0,width=1.0\columnwidth]{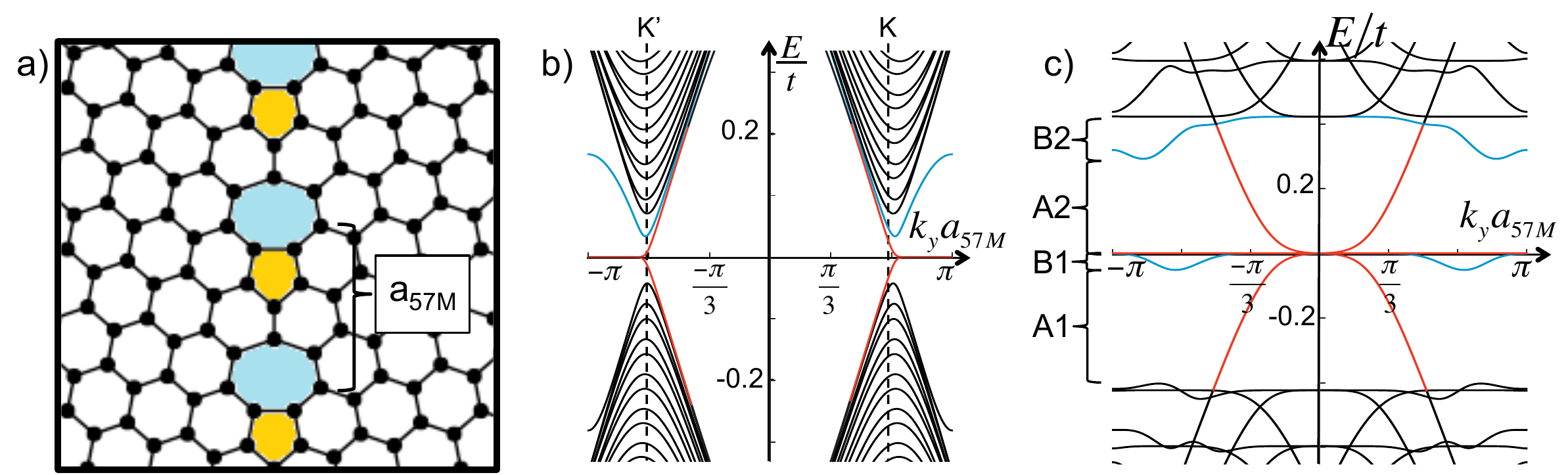}
  \caption{\label{Latt_Bands_Sup} {\bf a)} Graphene lattice containing a 5-7M twin grain boundary.  The GB structure consists of pentagon-hexagon dislocations that repeat with period $a_{57M}=\sqrt{7}a$. It is mirror symmetric across the GB line. {\bf b)} Band structure of a semi-infinite ribbon with a 5-7M t-GB along the infinite direction.  Bands are projected onto the 1D Brillouin zone of the GB. Surface (interfacial) bands are highlighted in red (blue).  The interfacial band is highly dispersive away from the K and K' points. {\bf c)} Band structure of the 5-7M ribbon in a perpendicular magnetic field with strength $B$ such that magnetic length $\ell_B = \sqrt{\hbar/eB} = 5.77a$.  Bands are projected onto the 1D Brillouin zone of the GB.  Surface (interfacial) bands are highlighted in red (blue).  Energy regions labelled A contain only edge-localized states while regions labelled B contain both edge and interfacial states. B regions correspond with the existence of ``bubbles" in the interfacial band that arise from hybridization of interfacial bands with the lowest LL.  The B regions cover a smaller energy range for the 5-7M t-GB as compared to the 5-7 or the 5-5-8 t-GBs because the interfacial band disperses strongly away from $E=0$ at zero field and therefore mixes comparatively little with the lowest Landau level. }
\end{figure}

The Kwant calculations were carried out for three grain boundary geometries: the 5-5-8 t-GB and the 5-7 t-GB referenced in the main text and an additional 5-7 t-GB with mirror symmetry, 5-7M, shown in Figure \ref{Latt_Bands_Sup}(a).  The 5-8-8 t-GB is of interest because it has been successfully fabricated in experiments \cite{S_Lahiri, S_Chen} and has a simple structure with no orientational mismatch between grains. The two 5-7 t-GB's, meanwhile, are known as ``large angle grain boundaries" (LAGB's) and are notable for their low formation energies \cite{S_YL1}. We display results for the 5-7M t-GB in this Supplemental Material as confirmation of the results in the main text and to compare it to the other LAGB. The band structure for the 5-7M system with $B = 0$ (Fig. \ref{Latt_Bands_Sup}b) shows two surface bands (red) and a highly dispersive interfacial band (blue).  In accordance with our results for 5-5-8 and 5-7, the 5-7M interfacial band mixes with the lowest Landau level (LLL) in a perpendicular B field (Fig. \ref{Latt_Bands_Sup}c), however the resulting ``bubbles" in the interfacial band are much smaller in energy because the zero field band disperses quickly away from $E=0$ and thus is less available for mixing with the LLL.

\begin{figure}
  \includegraphics[angle=0,width=0.9\columnwidth]{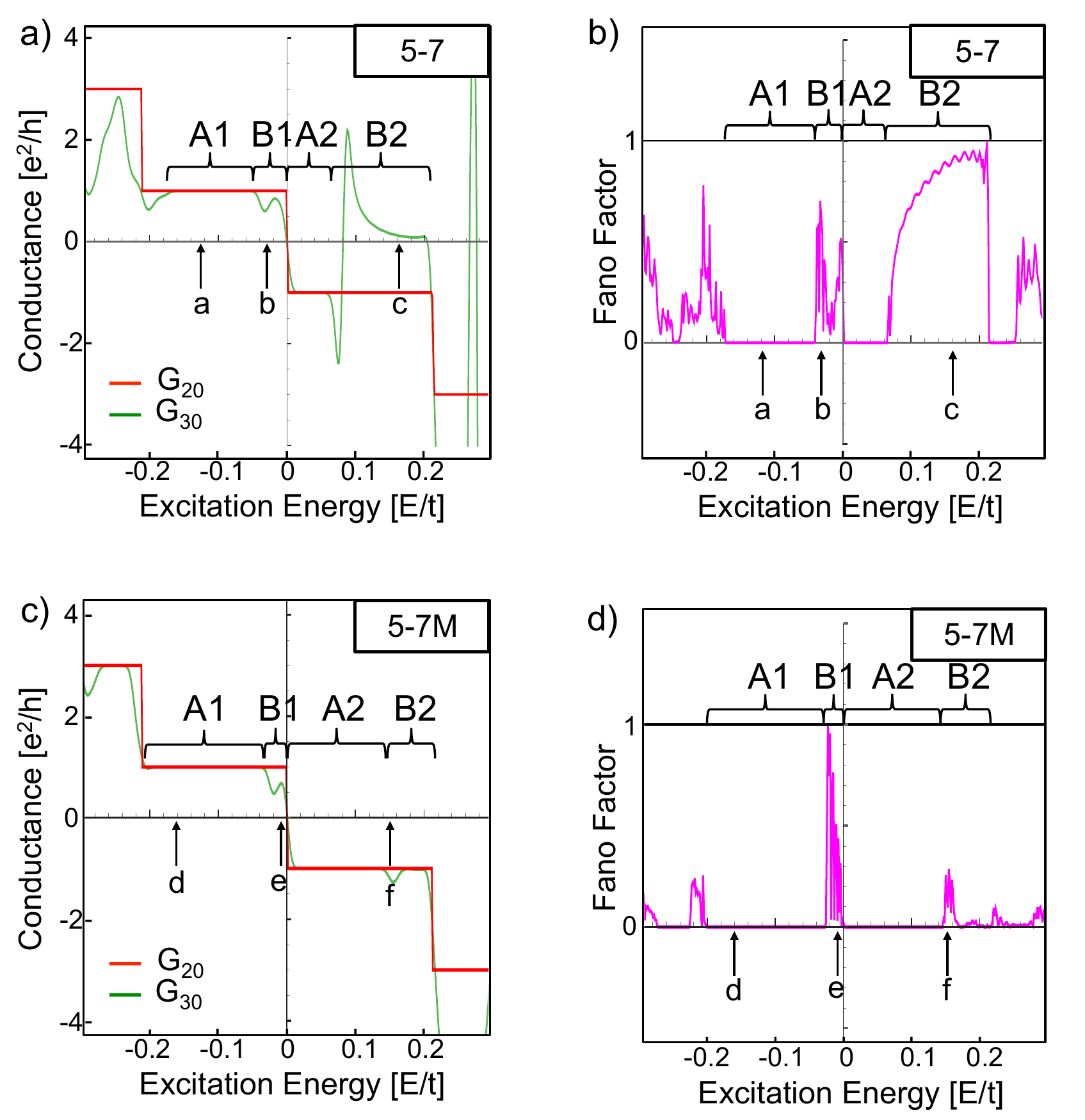}
  \caption{\label{Gs_Fs_Sup} Conductances for six-terminal Hall bars partitioned by {\bf a)} a 5-7 t-GB and {\bf c)} a 5-7M t-GB.  Red curves show the Hall conductance ($G_{20}$) measured between opposite leads on the same side of the GB.  Green curves show the cross-GB Hall conductance ($G_{30}$) measured between opposite leads on opposite sides of the GB.  The $G_{30}$ curve is Gaussian averaged with a peak width of $c=0.005E/t$.  $G_{20}$ exhibits conductance plateaus for all energies in both geometries.  In the energy regimes marked A in the primary Landau gap, $G_{30}=G_{20}$, which indicates that there is no longitudinal voltage drop across the GB.  Thus quantum Hall edge modes remain intact at these energies, as shown in Fig. \ref{States_Sup}a, d.  The A regions correspond with regions in the band structures where only surface bands exist.  In spectral regions marked B, dispersive edge and GB states coexist in energy due to hybridization of both types of states with the LLL state.  Coupling between the dispersive edge and interfacial states leads to a voltage drop across the GB, meaning $G_{30}\neq G_{20}$ in these energy regimes. The $G_{30}$ conductance maintains the Hall values at more energies in the 5-7M t-GB Hall bar than in other Hall bars studied because there is less hybridization of the interfacial band with the LLL and thus fewer energies where dispersive GB states and edge states coexist.  Fano factor measurements ($F$) on two-terminal devices partitioned by {\bf b)} a 5-7 t-GB and {\bf d)} a 5-7M t-GB confirm the conductance data.  $F=0$ over a larger portion of the first Landau level gap for the 5-7M system than for the 5-7 system, indicating that the quantum Hall edge states are preserved at more energies for this geometry than for the 5-7 geometry. Arrows labelled in all plots correspond to current maps shown in Fig. \ref{States_Sup}.}
\end{figure}

Consequently, when we carry out the scattering matrix calculations on the GB-partitioned Hall bar, we find that for the 5-7M geometry, the quantum Hall plateaus in the cross-GB conductance (green curve, Fig. \ref{Gs_Fs_Sup}c) are maintained nearly everywhere.  Compare this to the cross-GB conductance measured for the 5-7 t-GB, which has a less dispersive zero-field interfacial band and thus more hybridization of the LLL and GB states leading to stronger breakdown of the transverse conductance plateaus (Fig. \ref{Gs_Fs_Sup}a).  Accordingly, in the two terminal Fano factor ($F$) measurement for the 5-7 GB's, the 5-7 t-GB has a larger Fano factor over a wider range of energies than the 5-7M (Fig. \ref{Gs_Fs_Sup}b, d), indicating more energies at which edge states deflect into the grain boundary and back into the source lead.

\begin{figure}
  \includegraphics[angle=0,width=0.9\columnwidth]{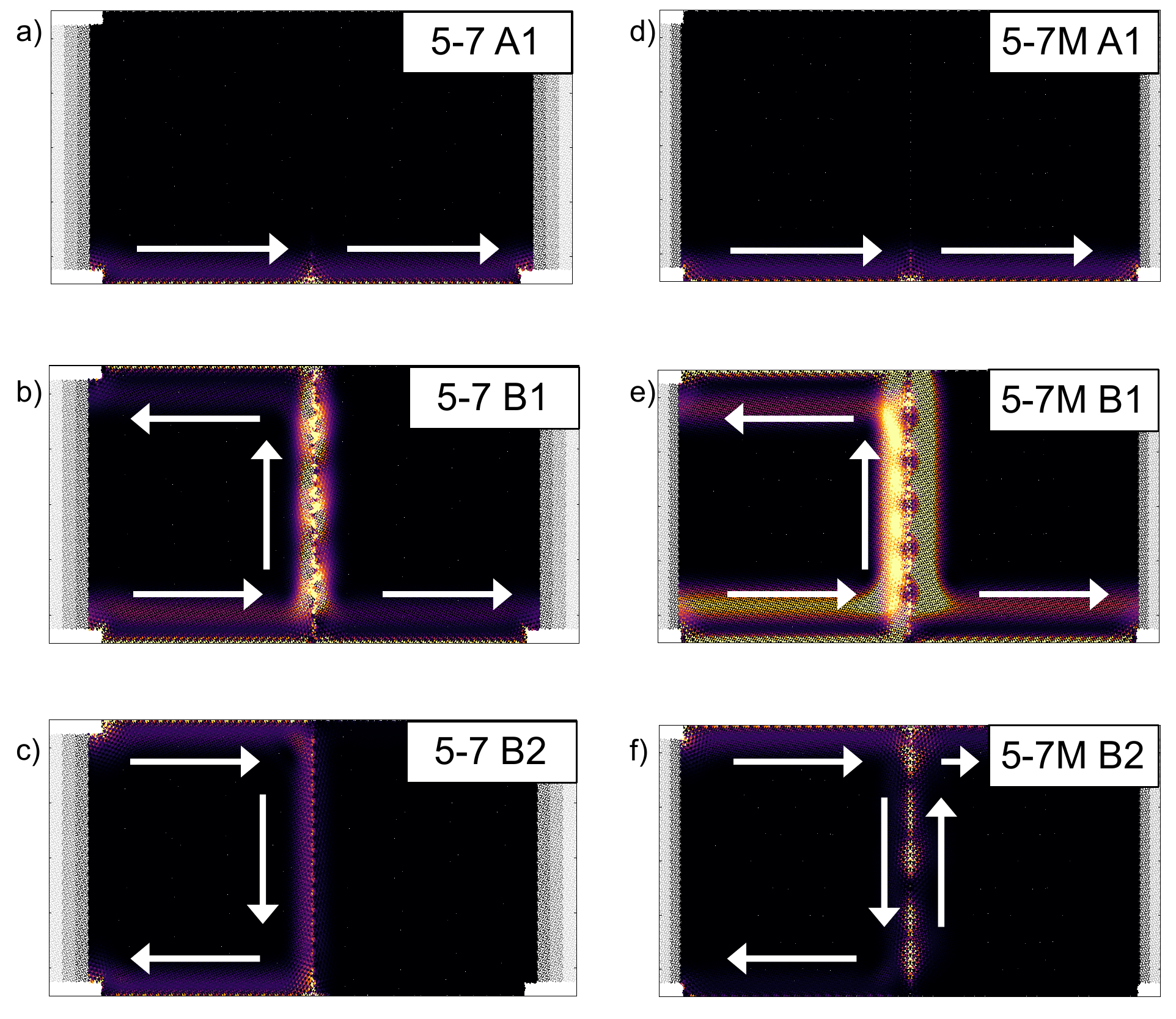}
  \caption{\label{States_Sup}  Density plots of current originating at source lead (left).  Panels {\bf a-c)} correspond to currents in two-terminal devices partitioned by a 5-7 t-GB, and panels {\bf d-f)} correspond to currents in two-terminal devices partitioned by a 5-7M t-GB.  Each panel is labelled by the geometry and energy region in which the current snapshot is taken.  The exact location within the energy regime is labelled by the arrows in Figure \ref{Gs_Fs_Sup}, with arrow {\bf a} corresponding to panel {\bf a)} of this plot, etc. Currents in the A energy regimes are the traditional quantum Hall edge states.  The B spectral regions have some degree of hybridization between interfacial states and the LLL.  In both 5-7 t-GB's, the B1 region primarily contains currents that are partially deflected back into the source via the GB and partially transmitted to the drain. The B2 spectral regions of the two systems differ.  The 5-7 B2 region contains currents that are almost completely deflected back into the source.  This corresponds to the B2 region in Fig. \ref{Gs_Fs_Sup}b where the Fano factor is nearly one.  The 5-7M t-GB has no such region of strong deflection at $E>0$.  Its B2 region sees only minimal hybridization of interfacial states with LL states.  This reduced hybridization leads to edge currents such as the one displayed in panel {\bf f)} that has some coupling with the non-chiral GB state but minimal scattering back into the source.}
\end{figure}

Maps of the current pathways (Fig. \ref{States_Sup}) confirm this result.  For both LAGB's, the currents in spectral region A1 (and A2, not shown) are traditional quantum Hall edge modes, and the currents in the B1 regions have at least some deflection into the GB and back into the source.  However, the B2 region of the 5-7 geometry has currents that nearly completely deflect back into the source via the GB while in 5-7M system, the B2 region is much smaller, and most of the current flows to the drain. Though there is some some backscattering in the GB, very little current is rerouted back into the source lead (Fig. \ref{States_Sup}f).

\end{document}